\documentclass[a4page]{revtex4}
\usepackage{graphicx,graphics} % Include figure files
\usepackage[dvips,bookmarks=false]{hyperref}

\usepackage{graphicx}
\usepackage{dcolumn}

\usepackage{eucal}
\usepackage[dvips]{epsfig}
\begin{document}
\title{Exchange coupling between two ferromagnetic electrodes
separated by a graphene nanoribbon}
\author{Alireza Saffarzadeh$^{1,2,}$}
\altaffiliation{E-mail: a-saffar@tehran.pnu.ac.ir}
\affiliation{$^1$Department of Physics,
Payame Noor University, Nejatollahi Street, 159995-7613 Tehran, Iran \\
$^2$Computational Physical Sciences Laboratory, Department of
Nano-Science, Institute for Research in Fundamental Sciences
(IPM), P.O. Box 19395-5531, Tehran, Iran}
%\date{\today}

\begin{abstract}
In this study, based on the self-energy method and the total
energy calculation, the indirect exchange coupling between two
semi-infinite ferromagnetic strips (FM electrodes) separated by
metallic graphene nanoribbons (GNRs) is investigated. In order to
form a FM/GNR/FM junction, a graphitic region of finite length is
coupled to the FM electrodes along graphitic zigzag or armchair
interfaces of width $N$. The numerical results show that, the
exchange coupling strength which can be obtained from the
difference between the total energies of electrons in the
ferromagnetic and antiferromagnetic couplings, has an oscillatory
behavior, and depends on the Fermi energy and the length of the
central region.
\end{abstract}
\maketitle

\section{Introduction}
Magnetic exchange interaction between ferromagnetic (FM) layers
separated by a nonmagnetic metal spacer has been investigated to a
large extent because of both possible applications exploiting the
giant magnetoresistance effect and the expectation of the way to
the production of nonvolatile computer memories, extremely
efficient magnetic sensors, and magnetic materials with enhanced
information storage capacity \cite{Hathaway,Heinrich}. Among
low-dimensional structures, graphene nanoribbons (GNRs) and carbon
nanotubes (CNTs) are potentially useful for magneto-transport
applications. For that purpose, GNRs and CNTs must interact with
magnetic foreign objects such as substrates, impurities, adsorbed
atoms, and nanoparticles \cite{Kirwan,Shenoy}. When the separation
between the magnetic objects increases, the direct exchange
interaction between magnetic moments decays abruptly since this
interaction requires a finite overlap between the wave functions
that surround the respective magnetic objects. Therefore, when the
moments are not too close together there is no direct overlap
between their wave functions. In such a case, the only way a
magnetic coupling can arise is if it is mediated by the conduction
electrons of the metallic host. Understanding the physics of this
so-called indirect exchange coupling between localized magnetic
moments mediated by the conduction electrons of metallic hosts
might provide clues to making magnetic carbon-based structures
which could be applicable in spintronic devices \cite{Costa}.

In the field of exchange coupling of magnetic trilayers or
multilayers, many methods and models have been employed, such as
the first-principles method \cite{Herman1}, the tight-binding
total energy calculation \cite{Hasegawa1}, the
Ruderman-Kittel-Kasuya-Yosida (RKKY) theory \cite{Bruno}, the
one-band tight-binding hole-confinement model \cite{Li}, the
free-electron model \cite{Barnas}, etc. In this article, based on
the single-band tight-binding approximation and the self-energy
method, we investigate the indirect exchange coupling between two
semi-infinite FM electrodes separated by metallic armchair GNRs
(AGNRs) and zigzag GNRs (ZGNRs).

\section{Model and formulation}
We consider a magnetic nanostructures consisting of two
semi-infinite FM nanostripes (electrodes) separated by a graphitic
region of length $L$ (in the $x$ direction) and width $N$ (in the
$y$ direction). Since we study the exchange coupling between two
FM electrodes, the electronic structure of the central part of the
junction (i.e. GNR) should be resolved in detail. It is therefore
reasonable to decompose the total Hamiltonian of the system as
$\hat{H}=\hat{H}_{L}+\hat{H}_{R}+\hat{H}_C+\hat{V}$. For
simplicity, we treat a square lattice structure for the FM
electrodes. The Hamiltonian for such electrodes is written within
the tight-binding approximation as
\begin{equation}\label{H}
\hat{H}_{\alpha}=\sum_{i_\alpha,\sigma}(\epsilon_{\alpha}-\mathbf{\sigma}\cdot
\mathbf{h}_{\alpha})\hat{c}_{i_\alpha,\sigma}^\dag\hat{c}_{i_\alpha,\sigma}-\sum_{<i_\alpha,
j_\alpha>}t_{i_\alpha,j_\alpha}\hat{c}_{i_\alpha,\sigma}^\dag\hat{c}_{j_\alpha,\sigma}\
,
\end{equation}
where $\hat{c}_{i_\alpha,\sigma}^\dag$
($\hat{c}_{i_\alpha,\sigma}$) creates (destroys) an electron with
spin $\sigma$ at site $i$ in electrode $\alpha$ (=L, R), and
$t_{i_\alpha,j_\alpha}=t$ is the hopping matrix element between
nearest-neighboring sites $i_\alpha$ and $j_\alpha$. Here,
$\epsilon_{\alpha}$ is the spin independent on-site energy and
will be set to zero, $-\mathbf{\sigma}\cdot \mathbf{h}_{\alpha}$
is the internal exchange energy with $\mathbf{h}_{\alpha}$
denoting the molecular field at site $i_\alpha$, and
$\mathbf{\sigma}$ being the conventional Pauli spin operator. On
the other hand, the Hamiltonian of GNR in the absence of FM
electrodes is expressed as
\begin{equation}\label{H}
\hat{H}_{C}=-\sum_{<i_C,
j_C>}t_{i_C,j_C}\hat{d}_{i_C,\sigma}^\dag\hat{d}_{j_C,\sigma}\ ,
\end{equation}
where $\hat{d}_{i_C,\sigma}^\dag$ ($\hat{d}_{i_C,\sigma}$) creates
(destroys) an electron with spin $\sigma$ at site $i$ of GNR.
Finally, $\hat{V}$ describes the coupling between the electrodes
and the GNR and takes the form $\hat{V}=-\sum_{i_\alpha,
j_C}t_{i_\alpha,j_C}(\hat{c}_{i_\alpha,\sigma}^\dag\hat{d}_{j_C,\sigma}+\textrm{H.c.})$.
The hopping elements $t_{i_C,j_C}$ between the $\pi$ orbitals of
the GNR, and also $t_{i_\alpha,j_C}$ between the lead orbitals and
the $\pi$ orbitals of the GNR are taken to be $t$; that is, the
hopping parameters are same in the three regions.

In this study we assume that the spin direction of the electron is
conserved in the propagating process through the GNR. Therefore,
there is no spin-flip process and the spin-dependent transport can
be decoupled into two spin channels: one for spin-up and the other
for spin-down. On the other hand, it is well known that the effect
of semi-infinite electrode $\alpha$ on the GNR can be described by
a (spin-dependent) self-energy matrix
$\hat{\Sigma}_{\alpha,\sigma}$ \cite{Datta}. Therefore, it is
reasonable to write $\hat{H}=\sum_\sigma \hat{H}_\sigma$, where
$\hat{H}_\sigma=\hat{H}_C+\hat{\Sigma}_{L,\sigma}+\hat{\Sigma}_{R,\sigma}$.
Now the spin-dependent Green's function of the GNR coupled to the
two FM electrodes is given as
\begin{equation}\label{G}
\hat{G}_\sigma(\epsilon)=[\epsilon \hat{1}
-\hat{H}_C-\hat{\Sigma}_{L,\sigma}(\epsilon)
-\hat{\Sigma}_{R,\sigma}(\epsilon)]^{-1}\ ,
\end{equation}
where the self-energy matrices contain the information of the
electronic structure of the FM electrodes and their coupling to
the GNR. These matrices can be expressed as
\begin{equation}\label{Self}
\hat{\Sigma}_{\alpha,\sigma}(\epsilon)=\hat{\tau}_{C,\alpha}\hat{g}_{\alpha,\sigma}(\epsilon)\hat{\tau}_{\alpha,C}\
,
\end{equation}
\begin{figure}[h]
\begin{minipage}{17pc}
\includegraphics[width=17pc]{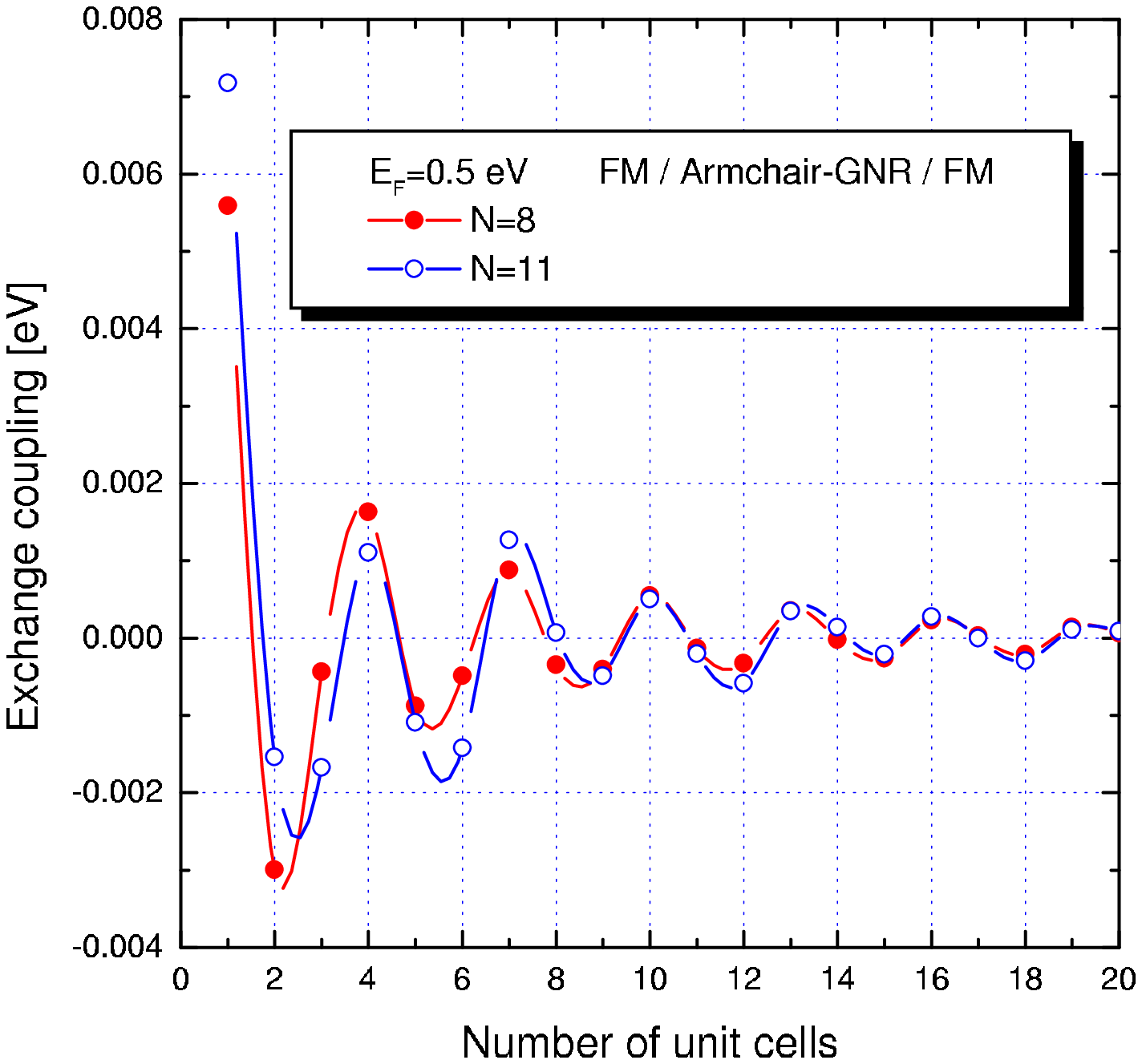}
\caption{$J$ as a function of AGNR length for $E_F$=0.5 eV.}
\end{minipage}\hspace{2pc}%
\begin{minipage}{17pc}
\includegraphics[width=17pc]{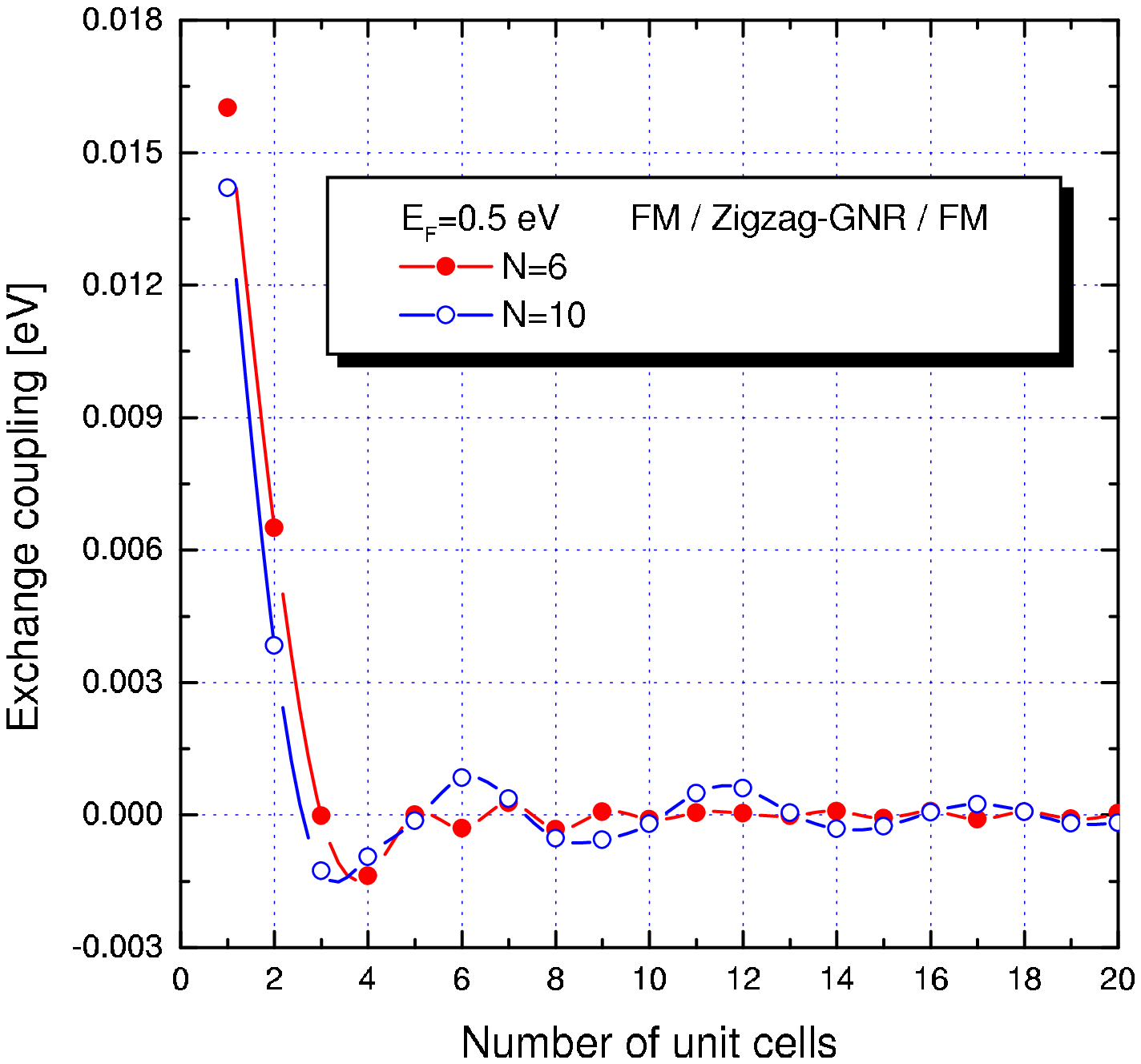}
\caption{$J$ as a function of ZGNR length for $E_F$=0.5 eV.}
\end{minipage}
\end{figure}

where $\hat{\tau}$ is the hopping matrix that couples the GNR to
the leads and is determined by the geometry of the GNR-lead bond.
$\hat{g}_{\alpha,\sigma}$ are the surface Green's functions of the
uncoupled leads i.e., the left and right semi-infinite magnetic
electrodes, and their matrix elements are given by
\begin{equation}\label{g}
g_{\alpha,\sigma}(i,j;z)=\frac{4}{N_x(N_y+1)}\sum_{l,k}\frac{\sin(kx_i)\sin(\frac{l\pi}{N_y+1}y_i)\sin(kx_j)\sin(\frac{l\pi}{N_y+1}y_j)}
{z-\epsilon_\alpha+\mathbf{\sigma}\cdot
\mathbf{h}_{\alpha}-2t[\cos(ka)+\cos(\frac{l\pi}{N_y+1})]}
\end{equation}
Here, $z=\epsilon+i\delta$, $1\leq l\leq N_y$,
$k\in[-\frac{\pi}{a},\frac{\pi}{a}]$ and $N_\beta$ with $\beta =x,
y$ is the number of lattice sites in the $\beta$ direction. In
order to calculate the total energy, we need the total density of
states per site for the electron of spin $\sigma$ (=$\uparrow$ or
$\downarrow$) which is given as $D_\sigma(\epsilon)=-\frac{1}{\pi
\mathcal{N}}\sum_i \mathrm{Im}[\hat{G}_\sigma(\epsilon)]_{ii}$
where $\mathcal{N}$ is the total number of carbon atoms in the
GNR. Thus, at $T=0$ K, the total energy of the electrons that
occupy the levels up to the Fermi energy $\epsilon_F$ in FM
(antiferromagnetic (AF)) configuration is given by
\begin{equation}
\mathcal{E}^{FM (AF)}=
\int_{-\infty}^{\epsilon_F}\epsilon[D_\uparrow^{FM
(AF)}(\epsilon)+D_\downarrow^{FM (AF)}(\epsilon)]d\epsilon\ .
\end{equation}

The indirect exchange coupling $J$, is defined as the total energy
difference between the FM and AF configurations of the system:
\begin{equation}
J=\mathcal{E}^{FM}-\mathcal{E}^{AF}\  .
\end{equation}
Based on the above formalism, we calculate $J$ for two different
junctions.
\begin{figure}[h]
\begin{minipage}{17pc}
\includegraphics[width=17pc]{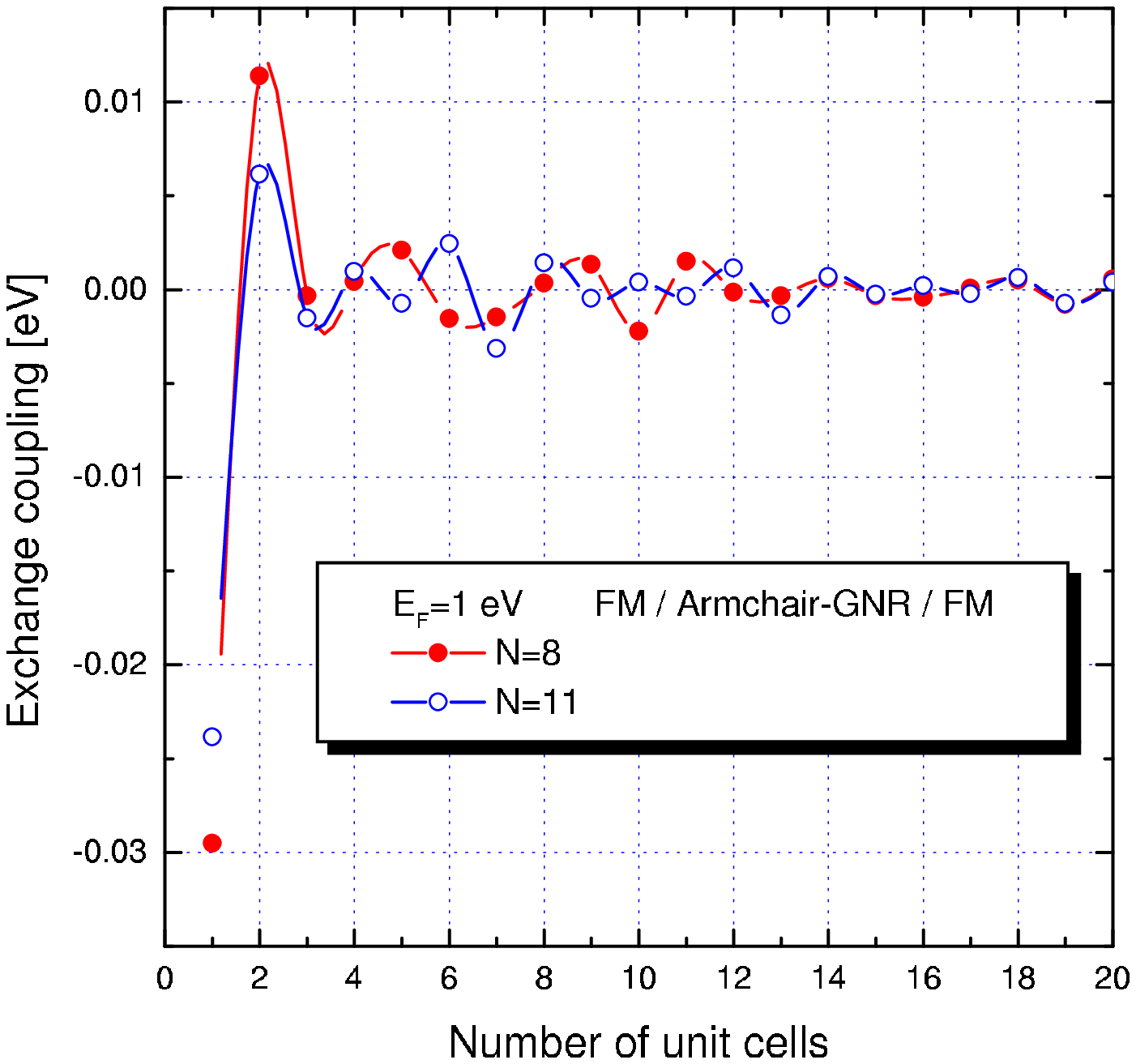}
%\caption{$J$ as a function of AGNR length for $E_F$=1 eV.}
\end{minipage}\hspace{2pc}%
\begin{minipage}{17pc}
\includegraphics[width=17pc]{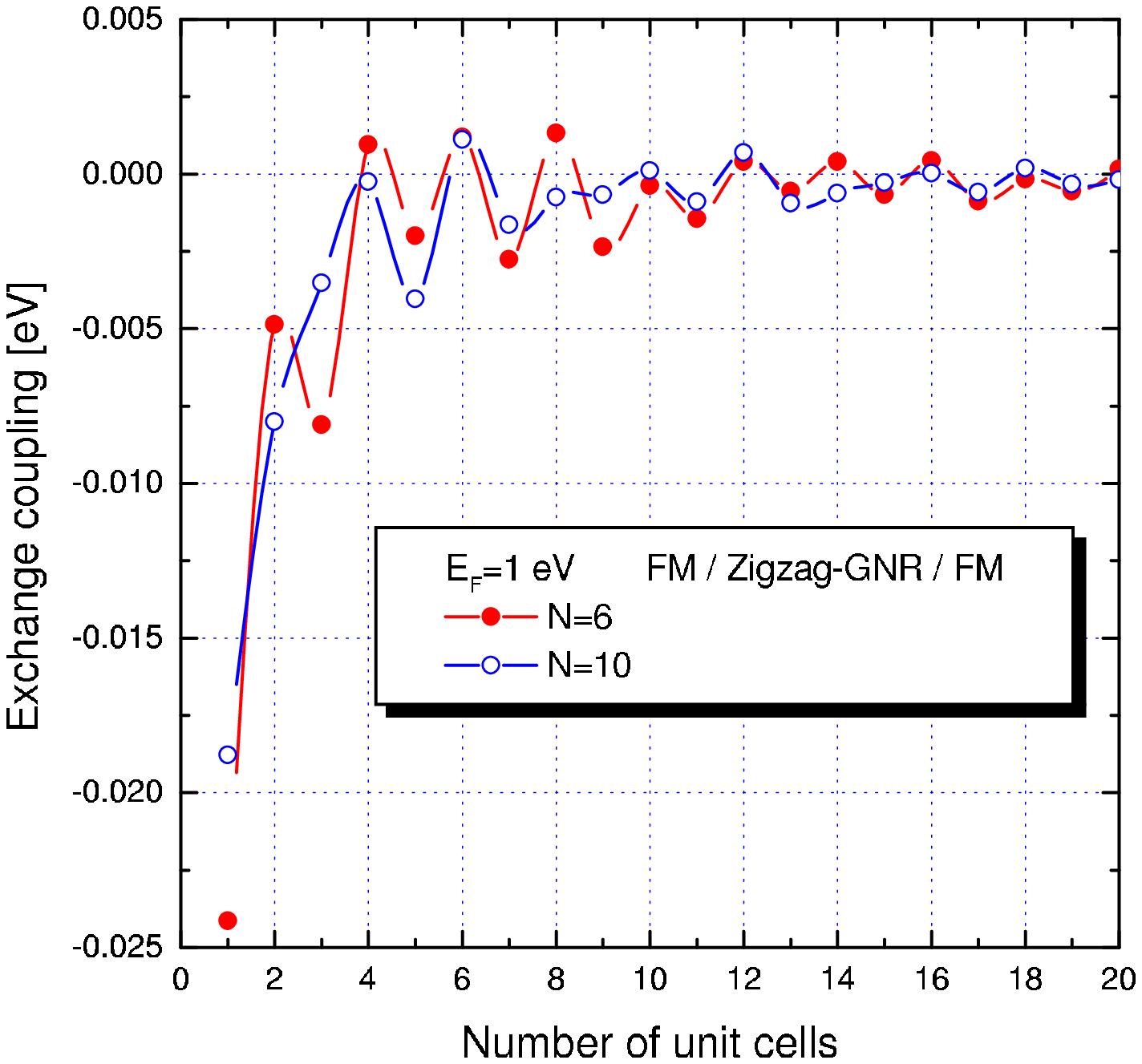}
%\caption{$J$ as a function of ZGNR length for $E_F$=1 eV.}
\end{minipage}
\end{figure}
\begin{figure}[h]
\begin{minipage}{17pc}
\includegraphics[width=17pc]{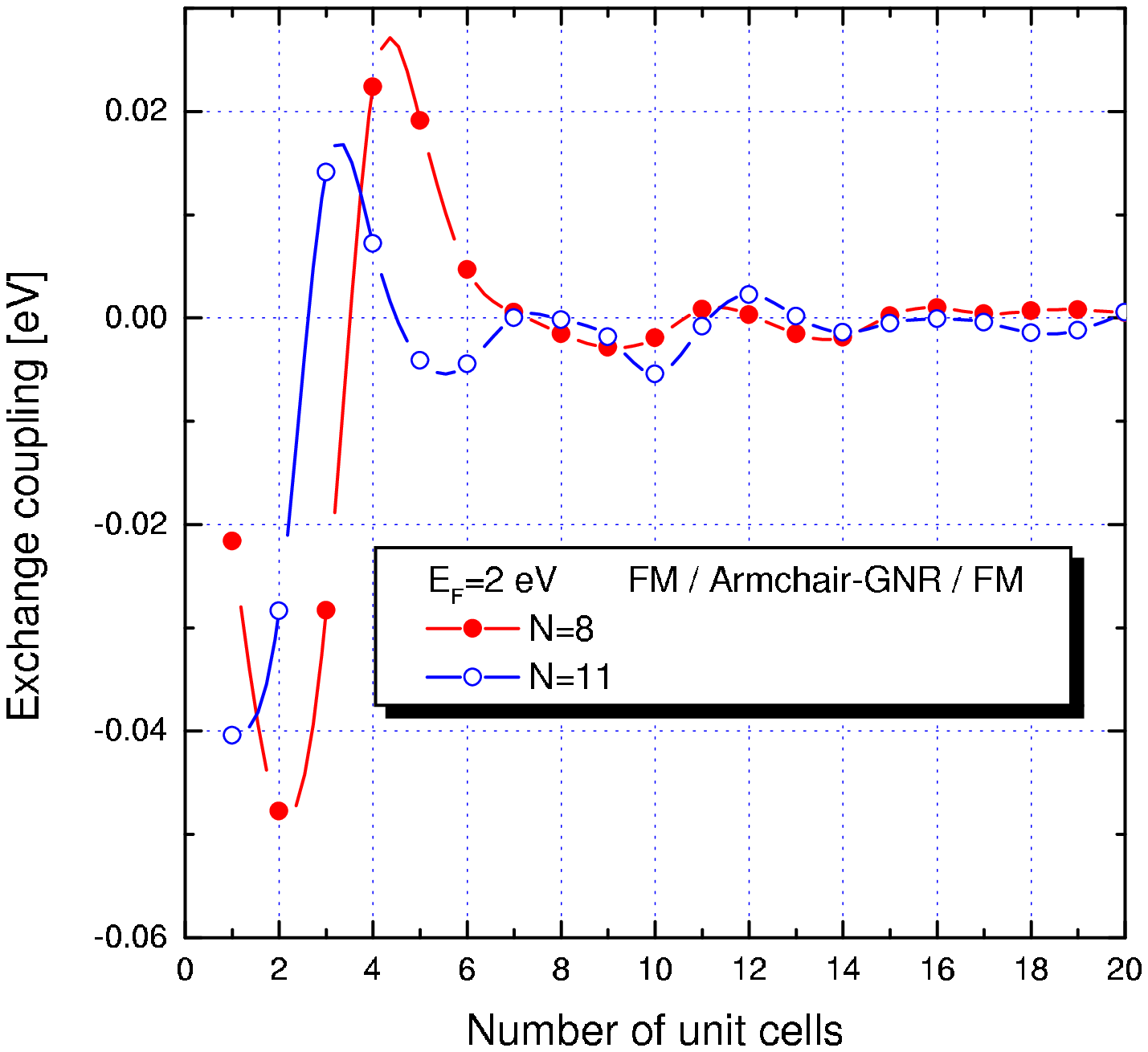}
\caption{$J$ as a function of AGNR length for $E_F$=1 eV and
$E_F$=2 eV.}
\end{minipage}\hspace{2pc}%
\begin{minipage}{17pc}
\includegraphics[width=17pc]{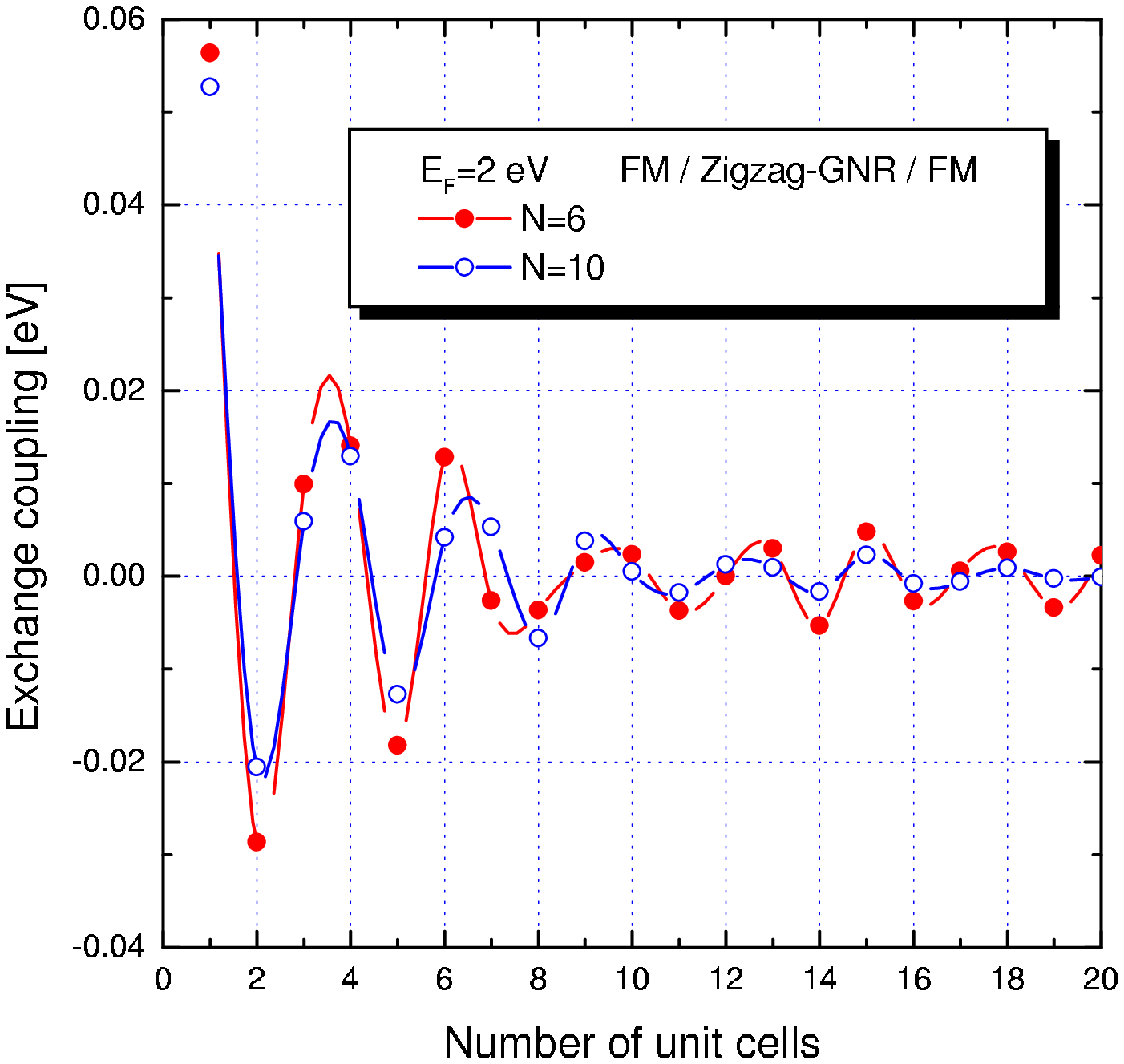}
\caption{$J$ as a function of ZGNR length for $E_F$=1 eV and
$E_F$=2 eV.}
\end{minipage}
\end{figure}

\section{Numerical results and discussions}
We have done the numerical calculations for the case that the
direction of magnetization in the left FM electrode is fixed in
the +$y$, while the magnetization in the right electrode is free
to be flipped into either the +$y$ or -$y$ direction. We set
$|\mathbf{h}_{\alpha}|$=1.5 eV and $t$=1 eV in the calculations.
Our results clearly show that $J$ oscillates with increasing the
length of graphitic region in both FM/AGNR/FM (Figs.1 and 3) and
FM/ZGNR/FM (Figs.2 and 4) junctions. It can be seen from the
figures that as the length of GNR increases, the oscillation
amplitudes of $J$ decrease. Also, the oscillatory phase changes by
varying the Fermi energy. We can ascribe the indirect coupling to
the interferences of electron waves in the metallic GNRs because
of spin-dependent reflections at the FM/GNR interfaces. Electron
states in the GNR can be propagative Bloch waves, but also
evanescent waves due to the finite length of the spacer.
Propagative states give rise to oscillatory contributions to the
coupling, while evanescent states yield damping terms.

Therefore, both kinds of states contribute to $J$. Their
importance depends essentially on the nature of the states at the
Fermi level. It is clear from the figures that, in most cases,
when the separation between two electrodes exceeds 15 unit cells,
the coupling between two magnetic electrodes becomes very small.
These results may be useful for understanding the strength of
exchange coupling between two magnetic impurities or adsorbed
atoms on GNRs and CNTs.

\end{document}